# Measurement of Quantum Weak Values of Photon Polarization

G. J. Pryde,[1,*] J. L. O'Brien,[1,*] A. G. White,[1,*] T. C. Ralph,[1] and H. M. Wiseman[2]

[1]*Centre for Quantum Computer Technology, Physics Department, The University of Queensland, Brisbane, QLD, 4072, Australia*
[2]*Centre for Quantum Computer Technology and Centre for Quantum Dynamics, Griffith University, Brisbane, QLD, 4111, Australia*

We experimentally determine weak values for a single photon's polarization, obtained via a weak measurement that employs a two-photon entangling operation, and postselection. The weak values cannot be explained by a semiclassical wave theory, due to the two-photon entanglement. We observe the variation in the size of the weak value with measurement strength, obtaining an average measurement of the $S_1$ Stokes parameter more than an order of magnitude outside of the operator's spectrum for the smallest measurement strengths.

It is commonly thought that the mean value of a quantum-mechanical measurement must be bounded by the extrema of a spectrum of eigenvalues, a consequence of statistical mathematics and the measurement postulate of quantum mechanics. However, there exist certain measurement outcomes for which this is not the case—these results are called weak values, since they arise as the outcomes of *weak measurements* on certain preselected and postselected quantum systems [1–11]. The canonical example of weak values is the *gedanken experiment* of Aharonov, Albert, and Vaidman [1], who described how it would be possible to use a weak measurement to measure (say) the $\sigma_z$ eigenvalue of a spin-1/2 particle, and determine an average value $\langle \sigma_z \rangle = 100$.

Weak values are an important and interesting phenomenon, because they assist us in understanding many couterintuitive results of quantum mechanics. For instance, weak values form a language by which we can resolve certain paradoxes and model strange quantum behavior. Important examples include: Hardy's paradox [12,13], in which two particles that always annihilate upon meeting are sometimes paradoxically measured after this annihilation event; the apparent superluminal transport of pulses in optical fibres displaying polarization mode dispersion [9]; apparently superluminal particles travelling in vacuum [8]; and quantifying momentum transfer in twin-slit "which-path" experiments [14–16]. Weak values are useful in simplifying calculations wherever a system is weakly coupled to a monitored environment [7,9]. They also are an example of a manifestly quantum phenomenon, in that the analysis of weak values can lead to negative (pseudo-) probabilities [12], an effect never observed in analogous classical measurements.

Here we present the first unambiguously quantum-mechanical experimental realization of weak values, where we use a nondeterministic entangling circuit to enable one single photon to make a weak measurement of the polarization of another, subject to certain preselections and postselections. Previous demonstrations of weak values using electromagnetic radiation [17–20] have used coherent states and weak measurements arising from the coupling of two degrees of freedom of the photon. They can thus be explained semiclassically using a wave equation derived from Maxwell's equations [21]. A cavity QED experiment [22] has been performed that was subsequently analyzed in terms of weak values [7], but the continuous spectrum precluded observations of anomalously large average values. By using two single photons, and realizing the weak measurement with a two-particle entangling operation, the weak values we measure (including extraspectral weak values) are not able to be described in semiclassical terms—a crucial result in the experimental verification and study of the phenomenon.

The observable we measure is the polarization of a single photon in the horizontal-vertical ($H-V$) basis, i.e., the quantum operator corresponding to the $S_1$ Stokes parameter [23], $\hat{S}_1 = |H\rangle\langle H| - |V\rangle\langle V|$, with expectation value $\langle \hat{S}_1 \rangle = \langle \mu | \hat{S}_1 | \mu \rangle$ for some state $|\mu\rangle$. According to the standard quantum formulation of measurement [24], $-1 \leq \langle \hat{S}_1 \rangle \leq 1$ for any single-photon polarization state. We will find that it is possible, using weak measurements, to obtain average values for $S_1$ far outside this range.

By analogy with the scheme of Aharonov, Albert, and Vaidman, we prepare the polarization of a single photon in the state

$$|\psi\rangle = \alpha|H\rangle + \beta|V\rangle, \qquad (1)$$

where $|\alpha|^2 + |\beta|^2 = 1$. Subsequently, we make a weak, nondestructive measurement on the photon's polarization in the $H-V$ basis. The weak measurement is made using a nondeterministic generalized photon polarization measurement device [25], which is deemed to have worked whenever a single photon is present at each of the *signal* and *meter* outputs. The generalized measurement device works by entangling the signal photon's polarization with that of a meter photon prepared in the state $\gamma|H\rangle + \bar{\gamma}|V\rangle$, before measuring the meter photon's polarization. Without loss of generality we choose $\gamma$ to be real; $\gamma^2 + \bar{\gamma}^2 = 1$. The experimental setup is shown in Fig. 1.

When operating with balanced modes [26], and after signal and meter photons interact but before either is

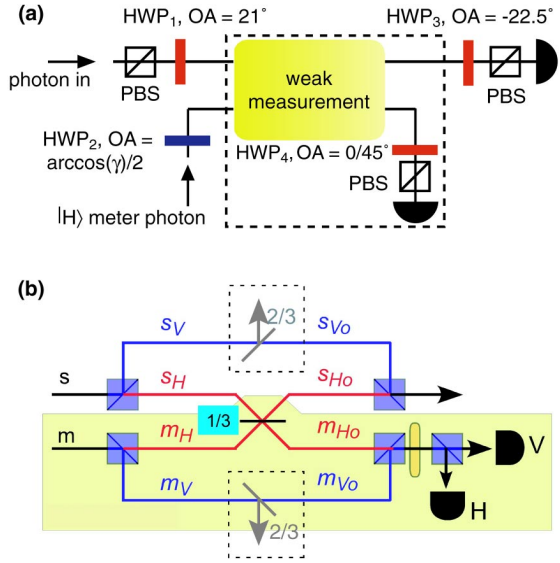

FIG. 1 (color online). (a) Conceptual representation of the experiment. A single photon is injected into the upper (signal) input port, where it is prepared in the state of Eq. (1) using a polarizing beam splitter (PBS) and half waveplate (HWP$_1$; OA = optic axis). A weak measurement of the polarization is made by interacting the photon with another (meter) photon in a weak measurement device, which operates via measurement-induced nonlinearity [25]. The interaction of the two photons can be controlled using HWP$_2$. The signal photon is then postselected in the state $|A\rangle = \frac{1}{\sqrt{2}}(|H\rangle - |V\rangle)$ using HWP$_3$, a PBS and a photon counter. A coincidence count flags successful postselection of the signal photon, and weak measurement with an outcome corresponding to the final meter waveplate (HWP$_4$) setting. The signal and meter photons are produced in pairs by spontaneous parametric downconversion from a beta-barium borate (BBO) crystal, pumped at 1.2 W at 351.1 nm by an Ar$^+$ laser. The mean coincidence rate without postselection on $|A\rangle$ was 44.6 s$^{-1}$ in 0.36 nm bandwidth; the postselection reduced this to 0.52 s$^{-1}$ across the measurement outcomes. Photon counting occurred over 100 s when making measurements in the absence of postselection on $|A\rangle$. This was increased to 1000 s when postselection on $|A\rangle$ was included. The coincidence window was 1 ns. We delivered the photons to the experiment through single mode optical fibers to provide Gaussian spatial modes for improved mode matching. (b) Conceptual representation of the weak measurement device [25,26]. The entangling operation occurs because of the nonclassical interference at the 1/3 beam splitter, and conditional on obtaining one and only one photon at the meter output.

measured, the state of the system in the two-qubit subspace corresponding to successful operation is

$$|\phi\rangle = (\alpha\gamma|H\rangle_s + \beta\bar{\gamma}|V\rangle_s)|H\rangle_m + (\alpha\bar{\gamma}|H\rangle_s + \beta\gamma|V\rangle_s)|V\rangle_m \quad (2)$$

where $s$, $m$ denote signal and meter photons, respectively. It follows that, with measurement of the meter photon in the $H - V$ basis, the weak measurement device implements a positive-operator-valued measurement (POVM) $\{\hat{\Pi}_H, \hat{\Pi}_V\}$ on the signal photon, with

$$\hat{\Pi}_i = \tfrac{1}{2}[\hat{1} + (\delta_{iH} - \delta_{iV})(2\gamma^2 - 1)\hat{S}_1], \quad (3)$$

where $\delta$ is the Kronecker delta. Equation (3) gives the measurement strength as $2\gamma^2 - 1$, which is set by the initial state of the meter photon. For a strong, projective measurement, $\gamma = 1$; weak measurement occurs when $\gamma$ is close to $1/\sqrt{2}$. A single weak measurement provides little information about the signal photon's polarization—the result is dominated by the randomness of measuring a meter state close to $(|H\rangle + |V\rangle)/\sqrt{2}$ in the $H - V$ basis. However, for a sufficiently large number of measurements on identically prepared photons, the average signal polarization can be recovered with arbitrary precision. The expectation value for $\hat{S}_1$ can be written in terms of probabilities of measuring $H$ or $V$ in the meter output:

$$\langle\hat{S}_1\rangle = \frac{\langle\psi|\hat{\Pi}_H|\psi\rangle - \langle\psi|\hat{\Pi}_V|\psi\rangle}{2\gamma^2 - 1} = \frac{P(H) - P(V)}{2\gamma^2 - 1}. \quad (4)$$

After making the weak measurement, we postselect the signal in a basis mutually unbiased with respect to $H - V$ (specifically, on the state $|A\rangle = \frac{1}{\sqrt{2}}|H\rangle - \frac{1}{\sqrt{2}}|V\rangle$). It is the selection of a subensemble of measurement results that can lead to the strange results of weak values. This leads to an expression for the postselected mean value:

$$_A\langle\hat{S}_1\rangle = \frac{P(H|A) - P(V|A)}{2\gamma^2 - 1}, \quad (5)$$

where the leading subscript represents the postselected state and where, for example, $P(H|A)$ denotes the probability of measuring $H$ in the meter output given that postselection on signal state $|A\rangle$ was successful. The general expression for the expected postselected value is then

$$_A\langle\hat{S}_1\rangle = \frac{\alpha^*\alpha - \beta^*\beta}{1 - 4\gamma\bar{\gamma}\mathrm{Re}[\alpha\beta]}. \quad (6)$$

Using Eq. (2), it can be shown that if $\gamma \rightarrow 1/\sqrt{2}$, then $_A\langle\hat{S}_1\rangle = \mathrm{Re}[(\alpha + \beta)/(\alpha - \beta)]$ so that when $\alpha - \beta \approx 0$, the weak value of $\hat{S}_1$ can be arbitrarily large. In practice, it is necessary to operate with nonzero measurement strength and postselection probability, so that a precise experimental value for $_A\langle\hat{S}_1\rangle$ can be obtained in a finite acquisition time. More detail on the theory of qubit weak values can be found in Ref. [27].

Postselected weak values are an important indicator of quantum behavior, since the bizarre results that can be obtained are not paralleled in the probabilities of analogous classical measurements. Large weak values arise from a quantum interference effect that results from the postselection of the signal photon state, which can be seen from the entangled state in Eq. (2). Consider the result when the meter photon is detected in the state $|H\rangle_m$, but no postselection is employed in the signal arm. The probability of this event is given by the expectation value of the projector

$\hat{1} \otimes |H\rangle_m \langle H|$, with the value $|\alpha\gamma|^2 + |\beta\bar{\gamma}|^2$ corresponding to the probability of measuring $H_s H_m$ plus the probability of measuring $V_s H_m$: the probabilities add, and there is no quantum interference. If we postselect on $|A\rangle$ in the signal arm, the probability of measuring $H$ in the meter, conditional on the postselection, is given by $(|\alpha\gamma - \beta\bar{\gamma}|^2)/(|\alpha\gamma - \beta\bar{\gamma}|^2 + |\alpha\bar{\gamma} - \beta\gamma|^2)$. It can be seen in the numerator that now the amplitudes add before squaring, allowing the possibility of a quantum interference effect. Combined with the similar expression for a $V$ measurement result, this leads to Eq. (6).

We measured the weak value of the single-photon polarization for a range of measurement strengths, with a nominal input state $|\psi\rangle = \cos(42°)|H\rangle + \sin(42°)|V\rangle \approx 0.743|H\rangle + 0.669|V\rangle$. In principle, the experimental value of $\gamma$ can be determined from the meter input waveplate settings. However, since the calculated values of $_A\langle \hat{S}_1 \rangle$ are very sensitive to $\gamma$, it is desirable to obtain the actual measurement strength from additional coincidence measurements, to deal with errors in the input waveplate setting and the remainder of the setup. The measurement strength is identical to the *knowledge* of the generalized measurement device, $K = P_{HH} + P_{VV} - P_{HV} - P_{VH} = 2\gamma^2 - 1$ (Ref. [25]), where, e.g., $P_{HV}$ is the probability of observing a horizontally and a vertically polarized photon at the signal and meter outputs of the device, respectively, and where these probabilities are measured with a signal input state $|D\rangle = (|H\rangle + |V\rangle)/\sqrt{2}$, and without postselecting the state $|A\rangle$. Because of Poissonian counting statistics in the measurement of $K$, the relative size of the error is quite large when $K$ is near zero.

The weak values for $\hat{S}_1$ were determined using Eq. (5) over a range of measurement strengths (Fig. 2). $P(H|A)$ and $P(V|A)$ were obtained from experimental coincidence measurements. For the smallest measurement strength, $K = 0.006$, we observed $_A\langle \hat{S}_1 \rangle = 47$, which is much larger than would be expected for a strong quantum-nondemolition measurement followed by postselection on $|A\rangle$, i.e., $_A\langle \hat{S}_1 \rangle = \alpha^*\alpha - \beta^*\beta \approx 0.1$, and also well outside the spectrum of $\hat{S}_1$.

The errors in $K = 2\gamma^2 - 1$ of approximately $\pm 0.015$ lead to substantial errors in the largest weak values, due to the form of Eq. (5). In fact, for the smallest measurement strengths, the uncertainty in $K$ encompasses $K = 0$, and the error in $|_A\langle \hat{S}_1 \rangle|$ is unbounded above. The triangles in Fig. 2 illustrate the "worst case" where each point is varied by $1\sigma$ in the value of $K$, in a direction that reduces the weak value. Even in this case, the smallest measurement strength yields a weak value of 19. In principle, the errors, which are all derived from Poissonian photon counting statistics, could be reduced arbitrarily by collecting larger samples of data. However, the low probability of the postselection, along with the very small correlation between the signal and meter photons, leads to very long collection times—a practical restriction on the size of the data set [28].

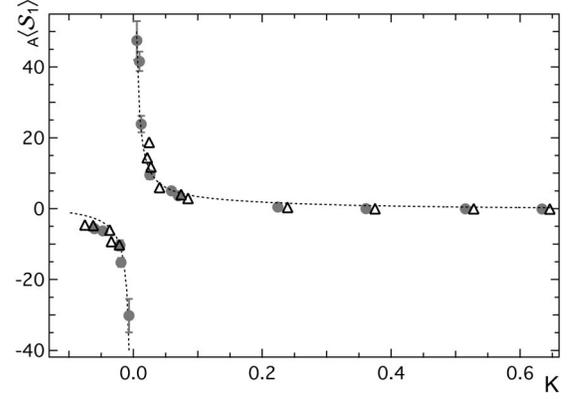

FIG. 2. Variation of experimentally observed weak values of $\hat{S}_1$ (circles) with measurement strength, $K = 2\gamma^2 - 1$. The error bars plotted arise from the effect of Poissonian counting statistics on $P(H|A)$ and $P(V|A)$. Bars not shown are smaller than the marker dimensions. In addition, errors in $K$ of approximately $\pm 0.015$ (over all values of $K$) lead to correlated errors in $_A\langle \hat{S}_1 \rangle$ via Eq. (5)—i.e., a displacement in $K$ due to error leads to a displacement in the weak value such that a given data point moves along a hyperbola. The triangles illustrate the worst case where each point is varied by $1\sigma$ in the value of $K$, in a direction that reduces the weak value. The dashed line indicates the predicted weak value, based on a model (with no free parameters) of the generalized measurement device obtained from quantum process tomography.

As the measurement strength is increased, we observe that the weak value of $S_1$ decreases until it is no longer greater than the strong value $|\alpha|^2 - |\beta|^2 \approx 0.1$. As noted in Ref. [25], the generalized measurement device does not exhibit perfect correlations between signal and meter due to imperfect mode matching. In the present case, this leads to a systematic offset in the weak value for large $K$, so that in fact it drops below this value.

The slight imperfections of the device mean that the theoretical weak value of Eq. (6), which assumes no mixture, does not completely describe the measurement. Instead, we determine the actual transfer matrix of the device—the process matrix—using quantum process tomography (in the manner of Ref. [29]). This provides an independent means of obtaining a parametric model of the postselected weak measurement process. As with Eq. (6), the expression obtained is parametrized by $\alpha$, $\beta$, $\gamma$, and $\bar{\gamma}$, although the slight mixture leads to a lengthier form [30]. The calculated curve for $_A\langle \hat{S}_1 \rangle$, plotted for our nominal input state, is shown in Fig. 2.

From a classical point of view, or even a typical quantum measurement point of view, it is quite strange that the measured expectation value of the $\hat{S}_1$ Stokes operator lies outside the interval $[-1, 1]$. The strangeness is perhaps more dramatic when we consider the results in terms of mean photon number. We can think of the expectation value of $\hat{S}_1$ as the difference in mode occupation between the two (spatially degenerate) $H$ and $V$ polarization modes. For instance, writing the strong measurement of $\hat{S}_1$ for a

single photon in state $|\psi\rangle$ in terms of mode occupation number:

$$\langle \hat{S}_1 \rangle = [\alpha^* \langle 1|_H \langle 0|_V + \beta^* \langle 0|_H \langle 1|_V](\hat{n}_H - \hat{n}_V)[\alpha|1\rangle_H|0\rangle_V$$
$$+ \beta|0\rangle_H|1\rangle_V] = \langle \hat{n}_H \rangle - \langle \hat{n}_V \rangle. \quad (7)$$

It follows that in the weak postselected case, ${}_\lambda\langle \hat{n}_H \rangle - {}_\lambda\langle \hat{n}_V \rangle = {}_\lambda\langle \hat{S}_1 \rangle$, for postselection on the state $|\lambda\rangle$. That is to say, we experimentally predict that conditional on preparing a single photon superposed across two polarization modes, and conditional on the measurement of $|A\rangle$ in the signal arm, there is a difference of as many as 47 photons between the two modes when we use the weakest generalized measurement. This seems nonsensical when we know that there was one signal input photon [31].

The resolution to this problem is that the weak values emerging from postselected weak measurements can be combined with the those from the complementary postselection to yield the expectation value of Eq. (4):

$$\langle \hat{S}_1 \rangle = {}_A\langle \hat{S} \rangle P(A) + {}_D\langle \hat{S} \rangle P(D) = \alpha\alpha^* - \beta\beta^*. \quad (8)$$

Again, the effects of mixture in the experimental process mean that this relationship does not exactly hold for the data—for instance, the weak measurement device slightly decoheres the signal photon, resulting in $P(A) = 0.012$ instead of the expected value, which is $\frac{1}{2} - \alpha\beta^* \approx 0.0027$ when $2\gamma^2 - 1 \approx 0$. Using the expected value for $P(A)$, and the measured weak value ${}_A\langle \hat{S}_1 \rangle = 47$, we obtain $\langle \hat{S}_1 \rangle = 0.25$. The standard error for $\langle \hat{S}_1 \rangle$ is bounded below by 0.10 and is unbounded above due to the error in ${}_A\langle \hat{S}_1 \rangle$. Using the expression obtained from the tomographic reconstruction, we can determine an accurate expectation value for $\hat{S}_1$ from the measured weak value, even in the presence of mixture. We obtain $\langle \hat{S}_1 \rangle = 0.08 \pm 0.03$ by this method. From the input settings, we expect $\langle \hat{S}_1 \rangle \approx 0.10$.

In conclusion, we have demonstrated a completely quantum realization of weak values. The weak measurement step relies on nonclassical interference between signal and meter photons, meaning that the results cannot be explained by Maxwell's equations alone. Using this technique, we observe expectation values of quantum observables far outside the range generally allowed by quantum measurement theory, including mean values of the single-photon $S_1$ Stokes parameter of up to 47.

We thank S. D. Bartlett for stimulating discussions. This work was supported by the Australian Research Council and the State of Queensland.